\documentclass[twocolumn,9pt,floatfix]{revtex4} 
\usepackage{epsfig}
\usepackage{amssymb}
\usepackage{subfigure}
\usepackage{color}
\usepackage{hyperref}
\usepackage{graphicx}
\usepackage{enumerate}
\usepackage[countmax]{subfloat}
\usepackage{MnSymbol}
\usepackage{accents}

\DeclareMathAlphabet{\mathpzc}{OT1}{pzc}{m}{it}

\def\beq{\begin{equation}}
\def\eeq{\end{equation}}

\def\bea{\arraycolsep .1em \begin{eqnarray}}
\def\eea{\end{eqnarray}}
\def\Tr{{\rm Tr}}

\def\g{\gamma}

\def\eq#1{(\ref{#1})}

\def\s0#1#2{\mbox{\small{$ \frac{#1}{#2} $}}}
\def\0#1#2{\frac{#1}{#2}}

\def\grgl{\:\hbox to -0.2pt{\lower2.5pt\hbox{$\sim$}\hss}{\raise3pt\hbox{$>$}}\:}
\def\klgl{\:\hbox to -0.2pt{\lower2.5pt\hbox{$\sim$}\hss}{\raise3pt\hbox{$<$}}\:}

\begin{document}

\title{Critical scaling in quantum gravity from the renormalisation group}
\author{Kevin Falls}
\affiliation{Instit\"{u}t f\"{u}r Theoretische Physik, Universit\"{a}t Heidelberg, Philosophenweg 12, 69120 Heidelberg, Germany}

\begin{abstract}
 The scaling behaviour of euclidean quantum gravity at an asymptotically safe critical point is studied by means of the exact renormalisation group.
Gauge independence is ensured via a specific parameterisation of metric fluctuations introduced in a recent paper.
Within a non-perturbative approximation the beta function for Newton's constant takes a simple form to all orders in $\hbar$. A UV fixed point is found to exist for $d\leq 7$ spacetime dimensions at which the critical scaling can be assessed.  The critical exponent for the Newton's constant $\nu$ is found to be regulator independent close to two dimensions. 
Applying Litim's optimisation criteria we find $\nu \approx 1/3$ in four spacetime dimensions. This value is in agreement with lattice studies supporting the existence of a second order phase transition between strongly and weakly coupled phases.

\end{abstract}

\date{March 24, 2015}

\maketitle

{\it Introduction}.--- An important open question in theoretical physics is whether a consistent renormalisable theory of gravity exists in $d=4$ dimensions. Asymptotic safety \cite{Weinberg:1980gg} offers one possible realisation of such a theory, whereby the ultra-violet (UV) limit is controlled by an interacting fixed point of the renormalisation group (RG).     
The essential idea of asymptotic safety is intimately related to critical phenomena in systems where a large number of collective degrees of freedom play a role. In particular, interacting fixed points generally occur where the physical system undergoes a second order phase transition \cite{Wilson:1973jj}. Thus, as in critical phenomena, a fixed point is characterised by a set of universal exponents. In quantum gravity such an exponent $\nu$ can be identified with the divergence of an RG invariant correlation length
\beq \label{xi}
\xi \propto \Lambda^{-1} \frac{1}{ |G_* - G_b|^{\nu}} \,,
\eeq
as the (dimensionless) bare Newton's constant $G_b$ is tuned to its critical value $G_*$ where $\Lambda$ is the UV energy scale (with $G_b \equiv G(\Lambda)= \Lambda^{d-2} G_{\Lambda}$ where $G_{\Lambda}$ denotes the  dimensionful bare coupling). Lacking any experimental means to measure such scaling behaviour, it is crucial that $\nu$ can be computed by complementary theoretical approaches.

Early progress was also made by exploiting the $\epsilon = d-2$ expansion whereby at two-loops, and up to order $\epsilon^2$, the value $1/\nu = \epsilon + \frac{3}{5} \epsilon^2$ has been found \cite{Aida:1996zn}. In four dimensional quantum gravity two main tools have been used to study critical behaviour, namely lattice models \cite{Hamber:1999nu,Ambjorn:2013tki} and the exact renormalisation group (ERG) \cite{Reuter:1996cp, Souma:1999at,  Litim:2003vp, Codello:2008vh, Benedetti:2009gn, Falls:2013bv}. On the lattice critical scaling can be  examined via quantities such as the mean curvature
\beq
\langle R \, \rangle \equiv \left\langle \int d^dx \sqrt{\g} \, R \, \right\rangle/V
\eeq
where $V$ is the $d$-dimensional volume and $R$ is the scalar curvature, which is integrated over spacetime, with all quantities in units of the lattice spacing $\ell = \Lambda^{-1}$.
One may then use the fixed point scaling properties of free energy $F(G_b) = - \log \mathcal{Z}(G_b)/V$ to infer that  \cite{Wegner:1972my}
\beq \label{Fscaling}
F(G_b) \sim  \xi^{-d}  \sim  |G_* - G_b|^{d \nu}  \,\,\, {\rm as}  \,\,\,\, G_b \to G_* \,,
\eeq 
where for gravity $\mathcal{Z}(G_b)= \int \mathcal Dg e^{ \frac{1}{16 \pi G_b}   \int d^dx \sqrt{\g} R + ... }$ is the partition function. The scaling behaviour of $\langle R \,\rangle$ can then be determined by differentiating with respect to the bare coupling $G_b$ to arrive at
$\langle R \, \rangle = 16 \pi G_b^2 \frac{\partial}{\partial G_{b}} F \sim  |G_* - G_b|^{ d \nu -1} \sim \xi^{1/\nu -d}$.
By studying the scaling behaviour of such quantities, for a lattice model based on Regge's simplicial formulation of gravity \cite{Regge:1961px},  the value \cite{Hamber:1999nu,Hamber}\,,
\beq \label{nuLat}
\nu_{\rm Lat} \approx 0.335(4) \,,
\eeq 
has been found in $d=4$ dimensions, where the error is due to statistical uncertainties. In addition arguments based on the non-local effective field equations \cite{Hamber:2006rq} imply that only $\nu= 1/3$ leads to consistent solutions.    
In this letter we will use the ERG to compute $\nu$ finding a value
\beq
\nu_{\rm ERG} \approx 1/3 \,.
\eeq
in $d=4$ for suitably optimised regulator schemes.
\newpage

{\it Gauge independence and the one-loop beta function}.--- 
One issue that must be resolved is the possible gauge dependence of $\nu$, which suggests that unphysical degrees of freedom contribute to the beta functions from which \eq{xi} can be obtained. 
In a recent paper a parameterisation of metric fluctuations was put forward for which gauge independence is ensured at the one-loop level \cite{Falls:2015qga}. 
The metric $\g_{\mu\nu}$ is given in terms of fluctuating fields around a background metric $g_{\mu\nu}$, taking the explicit form
\beq \label{para}
\g_{\mu\nu} =  \left(1 + \frac{\sigma}{2}\right)^{\frac{2}{d}}   g_{\mu\lambda} [e^{\hat{h}}]^{\lambda}_\nu\,  = g_{\mu\nu} + \hat{h}_{\mu\nu} + g_{\mu\nu} \frac{\sigma}{d} + ... \,,
\eeq
where $\hat{h}$ is a trace-free symmetric matrix with components $[\hat{h}]_{\mu\nu} \equiv \hat{h}_{\mu\nu}$ and $\sigma$ is a scalar field which parameterises the conformal fluctuations.
Gauge independence is achieved since the hessian of the bare action, expanded around an arbitrary Einstein background, does not involve terms proportional to the equations of motion. This property follows from \eq{para} since the volume element $\sqrt{\g} \propto 1 + \frac{\sigma}{2}$ is linear in $\sigma$. One then observes the cancelation of gauge fixed fields 
as well as the conformal fluctuations 
with the measure of the one-loop functional integral. This leads to the following beta function for the dimensionless Newton's constant in $d>2$ dimensions \cite{Falls:2015qga}
\beq \label{introbeta}
\beta_G =(d-2) G - \frac{2}{3}( 18-N_g) \, G^2 \,,\,\,\,\, N_g \equiv \frac{d(d-3)}{2}\,,
\eeq 
universal up to a rescaling of $G$, where $N_g$ denotes the polarisations of the graviton. For $d=2$ the contributions of the conformal fluctuations $\sigma$ are absent and one finds $\beta(G) =- \frac{2}{3}26\, G^2$ reflecting the $26$ dimensions of string theory.
 
 Given the beta function for Newton's constant the value of $\nu$ can be obtained from $\beta(G)$ by differentiating:
\beq \label{nudef}
1/\nu =- \left. \frac{\partial \beta_G}{\partial G} \right|_{G= G_*} = d-2 + {\rm quantum \,corrections}
\eeq
where $G_*$ is the values of $G$ at a fixed point for which $\beta(G_*)=0$.
One may then integrate the RG flow in the vicinity of $G_*$ whereby the correlation length \eq{xi} enters as an integration constant.
 At one-loop we simply reproduce the canonical scaling $1/\nu= d-2$ which follows from the dimensionality of the (inverse) Newton's constant.

{\it Flow equation}.---Here we shall study the flowing action $\Gamma_k[\varphi]$
which depends on the momentum scale $k$ and generalises the effective action $\Gamma$ to which it is equal in the limit $k \to 0$. 
An important property of $\Gamma_k[\varphi]$ is that it obeys the exact flow equation \cite{Wetterich:1992yh,Morris:1993qb} 
\beq \label{flow}
\partial_t \Gamma_k= \frac{1}{2} \rm STr   \frac{\partial_t \mathcal{R}_{k}}{ \Gamma^{(2)}_{k}+ \mathcal{R}_{k}}\,,
\eeq
where  $ t = \log(k/\Lambda)$ denotes the RG time.
In the case of gravity \cite{Reuter:1996cp} the fields $\varphi = \{\hat{h}_{\mu\nu}, \sigma, ... \} $ denotes the metric fluctuations fields as well ghost fields. The right side of flow equation is a super-trace involving the hessian of the action $\Gamma^{(2)}_{k}$ and an infra-red (IR) regulator $\mathcal{R}_{k}$. Utilising \eq{flow} will allow us to compute quantum corrections to the critical exponent \eq{nudef}.

The details of our calculation proceed along the lines of \cite{Falls:2015qga}, taking advantage of the parameterisation \eq{para}, but where we now consider the flowing action $\Gamma_k$, rather than the functional integral from which it can be derived \cite{Reuter:1996cp}.
In particular we consider an action $\Gamma_k$ of the Einstein Hilbert form,
\beq \label{action}
\Gamma_k = \int d^dx \sqrt{\g} \left(\lambda_k - \frac{R(\g_{\mu\nu})}{16 \pi  G_k} \right) +... \,,
\eeq
 plus gauge fixing, ghost and auxiliary terms from Jacobians in the functional measure. Here $G_k$ and $\lambda_k$ denote the $k$-dependent Newton's coupling and the vacuum energy respectively. Exploiting the parameterisation \eq{para} 
we can then find the RHS of \eq{flow} upon setting all fluctuating fields to zero. This constitutes a background field, or single metric, approximation to which we will confine ourselves, noting that methods that go beyond this \cite{Becker:2014qya, Codello:2013fpa, Christiansen:2014raa, Donkin:2012ud} could also be applied. Compared to the one-loop calculation the background field approximation implies an RG improvement leading to a beta function to all orders in $G$.   

To perform this RG improvement we assume that the hessian of ghosts and auxiliary fields depends on $G_k$ such that the unphysical degrees of freedom continue to cancel as in the one-loop approximation. These cancellation includes the conformal fluctuations and implicitly the non-propagating transverse-traceless fluctuations with the Jacobians.
The flow equation \eq{flow} then takes a form identical to the one-loop flow but with the bare Newton's constant $G_{\Lambda}$ replaced by the running coupling $G_k$  
\beq \label{flowphys}
\partial_t \Gamma_k = \frac{1}{2} \Tr \frac{\partial_t \mathcal{R}_{k,2 }}{ \Gamma^{(2)}_{k,2 } + \mathcal{R}_{k,2}} -\frac{1}{2} \Tr \frac{\partial_t \mathcal{R}_{k,1}}{ \Gamma^{(2)}_{k,1} + \mathcal{R}_{k,1}}\,, 
\eeq
where the first term is a trace over transverse-traceless fluctuations $h_{\mu\nu}^{\upvdash}$ and the second term is a trace over transverse vectors arising from the functional measure.
The hessians appearing in \eq{flowphys} are given by
\bea \label{Hessians}
\Gamma^{(2)}_{k,n} &=& (16 \pi G_k )^{-1} \Delta_{n}\,, \nonumber \\[2ex]
\Delta_2 \varphi_{\mu\nu} &=& \left( -\nabla^2 \varphi_{\mu\nu} - 2 R_{\mu}\,^\alpha\,_\nu\,^\beta \varphi_{\alpha \beta}\right) \,,
\\[2ex] \Delta_1 \varphi_\mu &=& \left(-\nabla^2 \delta_\mu^\nu- R_{\mu}\,^\nu \right) \varphi_\nu \nonumber
\eea
with $R_{\mu\nu}$, $R_{\mu \alpha\nu\beta} $ and $\nabla^2$ denoting the Ricci and Riemann curvatures and the laplacian respectively. 
The regulator $\mathcal{R}_{k}$ takes the form
\bea \label{RegChoice}
&&\Gamma_{k,n}^{(2)}(\Delta_n) + \mathcal{R}_{k,n} (\Delta_n) = \Gamma_{k,n}^{(2)}(\Delta_n \to P_n^2)\,, \\[2ex]
&&\, {\rm with } \,\,\,\,P_n^2 \equiv  \Delta_n + k^2 C(\Delta_n/k^2) \nonumber \,.
\eea
Here $C(z)$ is a dimensionless regulator function with the following properties
\beq \label{Cconditions}
C(1) =1\,, \,\,\, C(z \to \infty) \to 0 \,,\,\,\, 0 \neq C(0) = {\rm finite}\,,
\eeq  
but is otherwise an arbitrary monotonic function. The first condition is simply an arbitrary normalisation condition which removes the degeneracy of $C(z)$ under a rescaling $k^2 \to c k^2$ for which observables are invariant. The condition that the regulator vanishes for large momentum ensures that it is a Wilsonian regulator whereby the high energy modes are unsuppressed, while for low momentum the regulator behaves like a mass $\mathcal{R}_{k} \sim k^2$.

{\it Beta functions}.--- Using the early time heat kernel expansion for the operators \eq{Hessians}  one can evaluate the traces on the right side of \eq{flowphys}. Comparing terms up to linear order in the curvature 
we obtain the beta function $\partial_t G=\beta_G \equiv G (d-2 + \eta_G) $  for the dimensionless Newton's constant $G \equiv k^{d-2} G_k$ with
\beq \label{eta}
\eta_G =   \frac{ 2 \,(N_g-18)\, G \,\mathcal{I}_{d/2 -1}}{3\, (4 \pi)^{\frac{d-2}{2}} \Gamma \left(\frac{d}{2}-1\right) +   (N_g-18)\, G \, \tilde{\mathcal{I}}_{d/2 -1}   }\,,
\eeq
and the beta function  $\partial_t \lambda = \beta_\lambda$ for the dimensionless vacuum energy $\lambda= k^{-d} \lambda_k$, given by
\beq \label{betalambda}
\beta_\lambda = - d  \lambda   +  \frac{ N_g  \left( \mathcal{I}_{\frac{d}{2}}-\frac{1}{2}\eta _G
   \tilde{\mathcal{I}}_{\frac{d}{2}}\right)}{ (4 \pi)^{\frac{d}{2}}\Gamma \left(\frac{d}{2}\right)} \,.
\eeq
Here the regulator dependent integrals $\mathcal{I}_n$ and $\tilde{\mathcal{I}}_n$ take the form
\beq
\tilde{\mathcal{I}}_{n} = \int_0^\infty dz  \frac{  z^{n-1} C(z)}{ z +C(z)}\,,\,\, \mathcal{I}_{n} = \tilde{\mathcal{I}}_{n} - \int_0^\infty dz z^{n}  \frac{    C'(z)    }{ z +C(z)} \nonumber \,.
\eeq
We note that the quantum corrections to $\beta_G$ continue to be proportional to the universal factor $N_g-18$ found in the one-loop approximation.

{\it UV critical scaling}.--- Solving for $\beta(G_*) = 0$ one finds a fixed point for positive $G_*$ for $N_g <18$ and hence in $d \leq 7$ dimensions.
 At this fixed point the critical exponent for Newton's constant \eq{nudef} reads
\beq \label{critexpo}
1/\nu  = (d-2) + (d-2)^2 \frac{\tilde{\mathcal{I}}_{d/2-1}}{2\, \mathcal{I}_{d/2-1}}\,.
\eeq
This formula is in fact independent of the universal factor $\frac{2}{3}(N_g-18)$ which appears in $\beta(G)$ and arises from the heat kernel coefficients of $\Delta_1$ and $\Delta_2$.
Instead \eq{critexpo} owes its form to the general scaling properties of the flow equation. These properties are shared by a conformally reduced toy model \cite{Falls:2014zba} for which the critical exponent \eq{critexpo} was also found. We note that $\nu$ is real unlike the critical exponents found for gauge dependent RG flows which have been analysed in \cite{Nagy:2013hka}. 

If we take the limit $d \to 2$ both integrals appearing in \eq{critexpo} diverge with as $2/(d-2)$ and we may therefore determine the critical exponent to second order in the $\epsilon = d- 2$ expansion
\beq \label{thetaepsilon}
1/\nu = \epsilon +  \frac{1}{2} \epsilon^2 + \mathcal{O} \left(\epsilon^3\right) \,,
\eeq  
independently of the regulator function $C(z)$. Extrapolating to $\epsilon =2 $ this gives $\nu= 1/4$.
This result can be compared to the 2-loop result \cite{Aida:1996zn} which gives $1/\nu =  \epsilon +  \frac{3}{5} \epsilon^2 = 4.4$ for  $\epsilon =2 $. 

\begin{figure}[t]
\includegraphics[width=1.0\hsize]{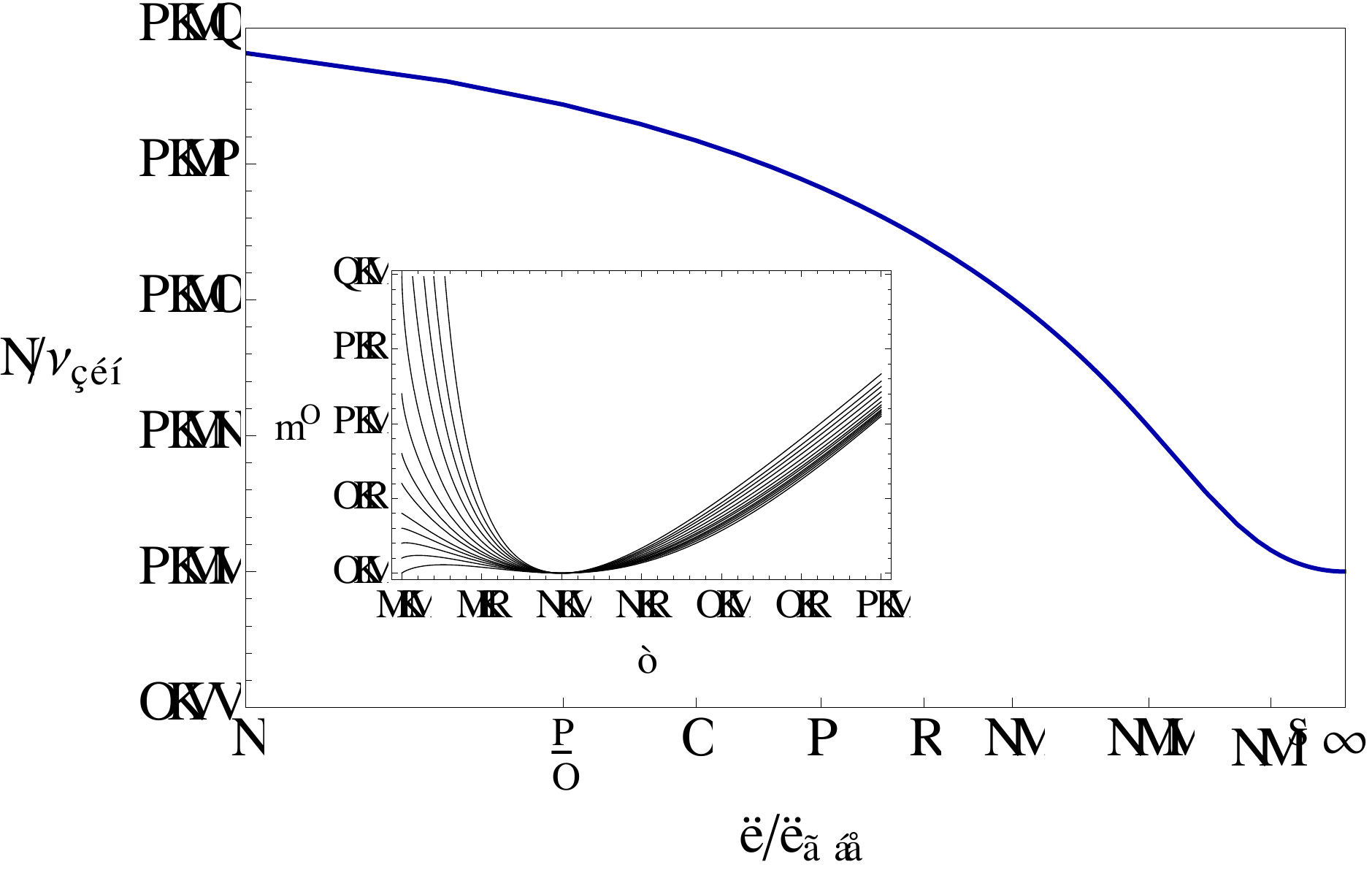}
\caption{\label{nuopt_p} The critical exponent $\nu$ plotted as a function of the regulator parameter $s$ which modifies the renormalisation scheme. The corresponding inverse propagators $P^2(z)$ are plotted in the inset with values of $s$ ranging from $s=s_{\rm min}$ to $s= \infty$ (bottom to top).}
\end{figure}

{\it Optimised scaling exponent in four dimensions.}--- 
Away from two dimensions we find that $\nu$ generally depends on the regulator. In what follows we will concentrate on the case $d=4$. In this case we have computed $\nu$ for a wide range of regulators $C(z)$, finding values in the range $1/4 < \nu < 1/2$. For example the entire range $1/4 < \nu < 1/2$ is found for the one parameter family $C(z) = \alpha \, \Theta(1-z)$ (where $\Theta(x)$ is the Heaviside theta function) for  which we find 
\beq
1/\nu = 2+ \frac{2 \alpha \log \left(\alpha^{-1}+1\right)}{\alpha \log
   \left(\alpha^{-1}+1\right)+\log (\alpha+1)} \,.
\eeq
for $0<\alpha< \infty$ (although $\alpha<1$ is not compatible with the normalisation  $C(1) = 1$).
Interestingly this corresponds to the entire range of values for which the phase transition is 
second order, as implied by \eq{Fscaling}, with the lower bound agreeing with the $\epsilon$-expansion \eq{thetaepsilon}. Furthermore we note that the value $\nu = 1/3$ is attained for $\alpha=1$.

A better determination of $\nu$, within the current approximation, is expected if we apply an optimisation criteria to the space of regulators $C(z)$.  
To this end we first note that there must be a non-zero gap in the inverse propagator
\beq \label{gap}
P^2_{\rm gap}[C]  \equiv P^2(z_{\rm min}) =  C(z_{\rm min}) + z_{\rm min}  \neq 0 \,,
\eeq
where $z_{\rm min}$ is the value of $z \geq 0$ for which $P^2(z)$ takes its global minimum. Generically we find that it is only in the limit where the gap vanishes $P^2_{\rm gap} \to 0$ that the value $\nu = 1/4$ is attained. To improve our estimate of $\nu$ we will apply Litim's
optimisation criteria \cite{Litim:2000ci} which is designed to improve the convergence of approximate solutions where only a finite number of operators (e.g. $\sqrt{\g}$ and $\sqrt{\g} R$) are retained. Specifically, we demand, in addition to \eq{Cconditions}, that the gap \eq{gap} is maximised under the regulator scheme (RS), which implies that
\beq \label{Copt}
P^2_{\rm gap}[C_{\rm opt}] =  \underaccent{\rm RS}{ \rm max} (P^2_{\rm gap}[C])= 2  \,,   
\eeq
where the specific value $\underaccent{\rm RS}{ \rm max} (P^2_{\rm gap})= 2$ is subject to our normalisation \eq{Cconditions}.
Any regulator $C = C_{\rm opt}$ which obeys \eq{Copt} is said to be optimised.
This criteria has been applied to the $3d$ Ising universality class in \cite{Litim:2010tt} leading to good convergence of the derivative expansion and a small systematic error in universal quantities. 
 A class of such optimised regulators is given by
\bea \label{CexpOpt}
C_{\rm opt}(z)&=& \left. s z^{b} \frac{1}{(1+s)^{z^b} -1} \right|_{b = b_{\rm opt}} \,, \nonumber \\[2ex]
{\rm where }\,\,\,  b_{\rm opt}(s) &\equiv& \frac{s}{(1+s)\log(1+s) - s}\,,
\eea
for values of $s$ $(b_{\rm opt})$  in the range $s_{\rm min} \leq s < \infty$ ($b_{\rm max} >  b_{\rm opt} > 0$), with $s_{\rm min}= 2.512$ $(b_{\rm max}=1.322)$,  for which \eq{Copt} is satisfied for the global minimum. We note that in the limit $s \to \infty $ we have $C_{\rm opt}(z) = 1/z$. The values of $1/\nu$ obtained for this class of regulators are plotted in fig.~\ref{nuopt_p} as a function of $s$. One observes that 
\beq
\nu_{\rm opt} \approx 1/3
\eeq
 for all permissible values of $s$ with a variation on the level of $\sim 1 \%$. For $s \to \infty$ we obtain $\nu = 1/3$ exactly.
 To check that this is not dependent on this class of regulators \eq{CexpOpt} one may also calculate $\nu$ for Litim's optimised regulator \cite{Litim:2001up} which has the form $(2 -z) \Theta(2-z)$ for which we again obtain $\nu = 1/3$. We have also evaluated $\nu$ with other optimised regulators and in each case found $\nu \approx 1/3$. We observe therefore that, although for generic regulators we find values $2< 1/\nu <4$, the application of the optimisation criteria \eq{Copt} predicts values $\nu \approx 1/3$ which agree quantitatively  with the lattice result \eq{nuLat}.

{\it Scaling of the free energy}.---
We now wish to make contact with the scaling arguments made in the {\it introduction} and in particular recover the scaling law for the free energy \eq{Fscaling}.  
This can be obtained from scaling of physical quantities at $k=0$ in the limit $G_b \to G_*$ where $G_b \equiv \Lambda^{d-2} G_{\Lambda}$. Provided $G_b < G_*$ the beta functions \eq{eta} and \eq{betalambda} for $G$ and $\lambda$ define a flow which leads to a classical regime at $k=0$ for which the dimensionful couplings flow to constants $G_{k\to 0} = G_N$ and $\lambda_{k \to 0} = \lambda_0$. These trajectories are then well defined for all $\Lambda \geq k \geq 0$. An individual trajectory is characterised by the dimensionless number $\tau_0 = 8 \pi G_N^{\frac{d}{d-2}}  \lambda_0$ obtained in the IR limit. We can then obtain the free energy from the effective action $\Gamma = \Gamma_0$
\beq \label{F}
F = \Gamma[\bar{g}_{\mu\nu}]/V = - \frac{2 \lambda_0}{d-2}  = - \frac{ \tau_0 }{4 \pi (d-2)  G_N^{\frac{d}{d-2}} } 
\eeq
where $\Gamma$ is evaluated on a solution to the equations of motion $\bar{g}_{\mu\nu}$.
 One may then integrate $\beta(G)$ between $G=G_b \equiv$ and $G =0$ to find the following scaling relation
\bea
G_N  &=& \Lambda^{2-d} \left. \varepsilon \exp \left[ (d-2) \int^{G_b}_\varepsilon dx \frac{1}{\beta(x)} \right] \right|_{\varepsilon \to 0, G_b \to G_*}   \nonumber \\[2ex]
 &=&  \Lambda^{2-d} G_* \left( \frac{G_* - G_b}{G_*}\right)^{-\nu(d-2)} +... \sim  \xi^{d-2}   \,,
\eea
Inserting this into \eq{F} we recover the scaling law \eq{Fscaling}.
In addition this scaling indicates that the Planck length $\ell_{\rm Pl} \equiv G_N^{\frac{1}{d-2}} \propto \xi$ emerges due to the existence of the fundamental length scale $\xi$.

{\it Discussion}.---In the lattice theory \cite{Hamber:1999nu} for which \eq{nuLat} was obtained the weak coupling phase 
$G_b<G_*$ gives rise to breached polymer phase which is usually understood as an effect of the conformal instability of the 
Euclidean action. 
Here we instead find a well behaved weak coupling phase with classical scaling. Indeed, we have implicitly Wick rotated the 
conformal factor  \cite{Mazur:1989by} resulting in the cancelation of conformal fluctuations with the functional measure. In turn on the lattice there is a strong coupling phase $G_b > G_*$ with spacetimes 
of negative curvature \cite{Beirl:1994wq,Hamber:1999nu} while for the beta function found here we run into a pole of $\eq{eta}$ for $G_b > G_*$. We observe therefore that currently the euclidean 
lattice theory only has access to a physical strong coupling phase whereas the ERG has access only to a weak phase. Since the conformal instability is due to 
the Euclidean nature of the path integral one may hope to get access to both phases via a suitable Wick rotation of the lattice theory. Such a 
rotation is offered by the causal dynamical triangulation (CDT) lattice theory \cite{Ambjorn:2013tki}, where a phase with classical scaling is observed \cite{Ambjorn:2004qm}. In turn one might hope that a strong coupling regime is accessible to the ERG once we go beyond the approximation used here e.g. by including higher order curvature invariants \cite{Codello:2008vh, Benedetti:2009gn, Falls:2013bv} or utilising vertex expansions \cite{Becker:2014qya, Codello:2013fpa, Christiansen:2014raa}.    

We conclude that the quantitative agreement between lattice and the continuum studies pressed here suggests that a second order phase transition exists between strongly and weakly coupled quantum gravity. This critical point could then provide a continuum limit for quantum gravity.

\section*{Acknowledgements}
The author would like to thank Jan Pawlowski and Daniel Litim for useful comments and Herbert Hamber for correspondence.

  \bibliography{myrefs,ASreferences2}

\begin{thebibliography}{30}
\expandafter\ifx\csname natexlab\endcsname\relax\def\natexlab#1{#1}\fi
\expandafter\ifx\csname bibnamefont\endcsname\relax
  \def\bibnamefont#1{#1}\fi
\expandafter\ifx\csname bibfnamefont\endcsname\relax
  \def\bibfnamefont#1{#1}\fi
\expandafter\ifx\csname citenamefont\endcsname\relax
  \def\citenamefont#1{#1}\fi
\expandafter\ifx\csname url\endcsname\relax
  \def\url#1{\texttt{#1}}\fi
\expandafter\ifx\csname urlprefix\endcsname\relax\def\urlprefix{URL }\fi
\providecommand{\bibinfo}[2]{#2}
\providecommand{\eprint}[2][]{\url{#2}}

\bibitem[{\citenamefont{Weinberg}(1979)}]{Weinberg:1980gg}
\bibinfo{author}{\bibfnamefont{S.}~\bibnamefont{Weinberg}}
  (\bibinfo{year}{1979}), \bibinfo{note}{in Hawking, S.W., Israel, W.: General
  Relativity}.

\bibitem[{\citenamefont{Wilson and Kogut}(1974)}]{Wilson:1973jj}
\bibinfo{author}{\bibfnamefont{K.~G.} \bibnamefont{Wilson}} \bibnamefont{and}
  \bibinfo{author}{\bibfnamefont{J.~B.} \bibnamefont{Kogut}},
  \bibinfo{journal}{Phys. Rept.} \textbf{\bibinfo{volume}{12}},
  \bibinfo{pages}{75} (\bibinfo{year}{1974}).

\bibitem[{\citenamefont{Aida and Kitazawa}(1997)}]{Aida:1996zn}
\bibinfo{author}{\bibfnamefont{T.}~\bibnamefont{Aida}} \bibnamefont{and}
  \bibinfo{author}{\bibfnamefont{Y.}~\bibnamefont{Kitazawa}},
  \bibinfo{journal}{Nucl.Phys.} \textbf{\bibinfo{volume}{B491}},
  \bibinfo{pages}{427} (\bibinfo{year}{1997}), \eprint{hep-th/9609077}.

\bibitem[{\citenamefont{Hamber}(2000)}]{Hamber:1999nu}
\bibinfo{author}{\bibfnamefont{H.~W.} \bibnamefont{Hamber}},
  \bibinfo{journal}{Phys. Rev.} \textbf{\bibinfo{volume}{D61}},
  \bibinfo{pages}{124008} (\bibinfo{year}{2000}), \eprint{hep-th/9912246}.

\bibitem[{\citenamefont{Ambjorn et~al.}(2013)\citenamefont{Ambjorn, Goerlich,
  Jurkiewicz, and Loll}}]{Ambjorn:2013tki}
\bibinfo{author}{\bibfnamefont{J.}~\bibnamefont{Ambjorn}},
  \bibinfo{author}{\bibfnamefont{A.}~\bibnamefont{Goerlich}},
  \bibinfo{author}{\bibfnamefont{J.}~\bibnamefont{Jurkiewicz}},
  \bibnamefont{and} \bibinfo{author}{\bibfnamefont{R.}~\bibnamefont{Loll}}
  (\bibinfo{year}{2013}), \eprint{1302.2173}.

\bibitem[{\citenamefont{Reuter}(1998)}]{Reuter:1996cp}
\bibinfo{author}{\bibfnamefont{M.}~\bibnamefont{Reuter}},
  \bibinfo{journal}{Phys. Rev.} \textbf{\bibinfo{volume}{D57}},
  \bibinfo{pages}{971} (\bibinfo{year}{1998}), \eprint{hep-th/9605030}.

\bibitem[{\citenamefont{Souma}(1999)}]{Souma:1999at}
\bibinfo{author}{\bibfnamefont{W.}~\bibnamefont{Souma}},
  \bibinfo{journal}{Prog. Theor. Phys.} \textbf{\bibinfo{volume}{102}},
  \bibinfo{pages}{181} (\bibinfo{year}{1999}), \eprint{hep-th/9907027}.

\bibitem[{\citenamefont{Litim}(2004)}]{Litim:2003vp}
\bibinfo{author}{\bibfnamefont{D.~F.} \bibnamefont{Litim}},
  \bibinfo{journal}{Phys.Rev.Lett.} \textbf{\bibinfo{volume}{92}},
  \bibinfo{pages}{201301} (\bibinfo{year}{2004}), \eprint{hep-th/0312114}.

\bibitem[{\citenamefont{Codello et~al.}(2009)\citenamefont{Codello, Percacci,
  and Rahmede}}]{Codello:2008vh}
\bibinfo{author}{\bibfnamefont{A.}~\bibnamefont{Codello}},
  \bibinfo{author}{\bibfnamefont{R.}~\bibnamefont{Percacci}}, \bibnamefont{and}
  \bibinfo{author}{\bibfnamefont{C.}~\bibnamefont{Rahmede}},
  \bibinfo{journal}{Annals Phys.} \textbf{\bibinfo{volume}{324}},
  \bibinfo{pages}{414} (\bibinfo{year}{2009}), \eprint{0805.2909}.

\bibitem[{\citenamefont{Benedetti et~al.}(2010)\citenamefont{Benedetti,
  Machado, and Saueressig}}]{Benedetti:2009gn}
\bibinfo{author}{\bibfnamefont{D.}~\bibnamefont{Benedetti}},
  \bibinfo{author}{\bibfnamefont{P.~F.} \bibnamefont{Machado}},
  \bibnamefont{and}
  \bibinfo{author}{\bibfnamefont{F.}~\bibnamefont{Saueressig}},
  \bibinfo{journal}{Nucl. Phys.} \textbf{\bibinfo{volume}{B824}},
  \bibinfo{pages}{168} (\bibinfo{year}{2010}), \eprint{0902.4630}.

\bibitem[{\citenamefont{Falls et~al.}(2013)\citenamefont{Falls, Litim,
  Nikolakopoulos, and Rahmede}}]{Falls:2013bv}
\bibinfo{author}{\bibfnamefont{K.}~\bibnamefont{Falls}},
  \bibinfo{author}{\bibfnamefont{D.}~\bibnamefont{Litim}},
  \bibinfo{author}{\bibfnamefont{K.}~\bibnamefont{Nikolakopoulos}},
  \bibnamefont{and} \bibinfo{author}{\bibfnamefont{C.}~\bibnamefont{Rahmede}}
  (\bibinfo{year}{2013}), \eprint{hep-th/1301.4191}.

\bibitem[{\citenamefont{Wegner}(1972)}]{Wegner:1972my}
\bibinfo{author}{\bibfnamefont{F.~J.} \bibnamefont{Wegner}},
  \bibinfo{journal}{Phys.Rev.} \textbf{\bibinfo{volume}{B5}},
  \bibinfo{pages}{4529} (\bibinfo{year}{1972}).

\bibitem[{\citenamefont{Regge}(1961)}]{Regge:1961px}
\bibinfo{author}{\bibfnamefont{T.}~\bibnamefont{Regge}},
  \bibinfo{journal}{Nuovo Cim.} \textbf{\bibinfo{volume}{19}},
  \bibinfo{pages}{558} (\bibinfo{year}{1961}).

\bibitem[{\citenamefont{Hamber}(2015)}]{Hamber}
\bibinfo{author}{\bibfnamefont{H.}~\bibnamefont{Hamber}}, \bibinfo{journal}{to
  be published}  (\bibinfo{year}{2015}).

\bibitem[{\citenamefont{Hamber and Williams}(2006)}]{Hamber:2006rq}
\bibinfo{author}{\bibfnamefont{H.~W.} \bibnamefont{Hamber}} \bibnamefont{and}
  \bibinfo{author}{\bibfnamefont{R.~M.} \bibnamefont{Williams}},
  \bibinfo{journal}{Phys.Lett.} \textbf{\bibinfo{volume}{B643}},
  \bibinfo{pages}{228} (\bibinfo{year}{2006}), \eprint{gr-qc/0607131}.

\bibitem[{\citenamefont{Falls}(2015)}]{Falls:2015qga}
\bibinfo{author}{\bibfnamefont{K.}~\bibnamefont{Falls}} (\bibinfo{year}{2015}),
  \eprint{1501.05331}.

\bibitem[{\citenamefont{Wetterich}(1993)}]{Wetterich:1992yh}
\bibinfo{author}{\bibfnamefont{C.}~\bibnamefont{Wetterich}},
  \bibinfo{journal}{Phys. Lett.} \textbf{\bibinfo{volume}{B301}},
  \bibinfo{pages}{90} (\bibinfo{year}{1993}).

\bibitem[{\citenamefont{Morris}(1994)}]{Morris:1993qb}
\bibinfo{author}{\bibfnamefont{T.~R.} \bibnamefont{Morris}},
  \bibinfo{journal}{Int.J.Mod.Phys.} \textbf{\bibinfo{volume}{A9}},
  \bibinfo{pages}{2411} (\bibinfo{year}{1994}), \eprint{hep-ph/9308265}.

\bibitem[{\citenamefont{Becker and Reuter}(2014)}]{Becker:2014qya}
\bibinfo{author}{\bibfnamefont{D.}~\bibnamefont{Becker}} \bibnamefont{and}
  \bibinfo{author}{\bibfnamefont{M.}~\bibnamefont{Reuter}},
  \bibinfo{journal}{Annals Phys.} \textbf{\bibinfo{volume}{350}},
  \bibinfo{pages}{225} (\bibinfo{year}{2014}), \eprint{1404.4537}.

\bibitem[{\citenamefont{Codello et~al.}(2014)\citenamefont{Codello, D'Odorico,
  and Pagani}}]{Codello:2013fpa}
\bibinfo{author}{\bibfnamefont{A.}~\bibnamefont{Codello}},
  \bibinfo{author}{\bibfnamefont{G.}~\bibnamefont{D'Odorico}},
  \bibnamefont{and} \bibinfo{author}{\bibfnamefont{C.}~\bibnamefont{Pagani}},
  \bibinfo{journal}{Phys.Rev.} \textbf{\bibinfo{volume}{D89}},
  \bibinfo{pages}{081701} (\bibinfo{year}{2014}), \eprint{1304.4777}.

\bibitem[{\citenamefont{Christiansen et~al.}(2014)\citenamefont{Christiansen,
  Knorr, Pawlowski, and Rodigast}}]{Christiansen:2014raa}
\bibinfo{author}{\bibfnamefont{N.}~\bibnamefont{Christiansen}},
  \bibinfo{author}{\bibfnamefont{B.}~\bibnamefont{Knorr}},
  \bibinfo{author}{\bibfnamefont{J.~M.} \bibnamefont{Pawlowski}},
  \bibnamefont{and} \bibinfo{author}{\bibfnamefont{A.}~\bibnamefont{Rodigast}}
  (\bibinfo{year}{2014}), \eprint{1403.1232}.

\bibitem[{\citenamefont{Donkin and Pawlowski}(2012)}]{Donkin:2012ud}
\bibinfo{author}{\bibfnamefont{I.}~\bibnamefont{Donkin}} \bibnamefont{and}
  \bibinfo{author}{\bibfnamefont{J.~M.} \bibnamefont{Pawlowski}}
  (\bibinfo{year}{2012}), \eprint{1203.4207}.

\bibitem[{\citenamefont{Falls}(2014)}]{Falls:2014zba}
\bibinfo{author}{\bibfnamefont{K.}~\bibnamefont{Falls}} (\bibinfo{year}{2014}),
  \eprint{1408.0276}.

\bibitem[{\citenamefont{Nagy et~al.}(2013)\citenamefont{Nagy, Fazekas, Juhasz,
  and Sailer}}]{Nagy:2013hka}
\bibinfo{author}{\bibfnamefont{S.}~\bibnamefont{Nagy}},
  \bibinfo{author}{\bibfnamefont{B.}~\bibnamefont{Fazekas}},
  \bibinfo{author}{\bibfnamefont{L.}~\bibnamefont{Juhasz}}, \bibnamefont{and}
  \bibinfo{author}{\bibfnamefont{K.}~\bibnamefont{Sailer}},
  \bibinfo{journal}{Phys.Rev.} \textbf{\bibinfo{volume}{D88}},
  \bibinfo{pages}{116010} (\bibinfo{year}{2013}), \eprint{1307.0765}.

\bibitem[{\citenamefont{Litim}(2000)}]{Litim:2000ci}
\bibinfo{author}{\bibfnamefont{D.~F.} \bibnamefont{Litim}},
  \bibinfo{journal}{Phys.Lett.} \textbf{\bibinfo{volume}{B486}},
  \bibinfo{pages}{92} (\bibinfo{year}{2000}), \eprint{hep-th/0005245}.

\bibitem[{\citenamefont{Litim and Zappala}(2011)}]{Litim:2010tt}
\bibinfo{author}{\bibfnamefont{D.~F.} \bibnamefont{Litim}} \bibnamefont{and}
  \bibinfo{author}{\bibfnamefont{D.}~\bibnamefont{Zappala}},
  \bibinfo{journal}{Phys.Rev.} \textbf{\bibinfo{volume}{D83}},
  \bibinfo{pages}{085009} (\bibinfo{year}{2011}), \eprint{1009.1948}.

\bibitem[{\citenamefont{Litim}(2001)}]{Litim:2001up}
\bibinfo{author}{\bibfnamefont{D.~F.} \bibnamefont{Litim}},
  \bibinfo{journal}{Phys.Rev.} \textbf{\bibinfo{volume}{D64}},
  \bibinfo{pages}{105007} (\bibinfo{year}{2001}), \eprint{hep-th/0103195}.

\bibitem[{\citenamefont{Mazur and Mottola}(1990)}]{Mazur:1989by}
\bibinfo{author}{\bibfnamefont{P.~O.} \bibnamefont{Mazur}} \bibnamefont{and}
  \bibinfo{author}{\bibfnamefont{E.}~\bibnamefont{Mottola}},
  \bibinfo{journal}{Nucl.Phys.} \textbf{\bibinfo{volume}{B341}},
  \bibinfo{pages}{187} (\bibinfo{year}{1990}).

\bibitem[{\citenamefont{Beirl et~al.}(1994)\citenamefont{Beirl, Gerstenmayer,
  Markum, and Riedler}}]{Beirl:1994wq}
\bibinfo{author}{\bibfnamefont{W.}~\bibnamefont{Beirl}},
  \bibinfo{author}{\bibfnamefont{E.}~\bibnamefont{Gerstenmayer}},
  \bibinfo{author}{\bibfnamefont{H.}~\bibnamefont{Markum}}, \bibnamefont{and}
  \bibinfo{author}{\bibfnamefont{J.}~\bibnamefont{Riedler}},
  \bibinfo{journal}{Phys.Rev.} \textbf{\bibinfo{volume}{D49}},
  \bibinfo{pages}{5231} (\bibinfo{year}{1994}), \eprint{hep-lat/9402002}.

\bibitem[{\citenamefont{Ambjorn et~al.}(2004)\citenamefont{Ambjorn, Jurkiewicz,
  and Loll}}]{Ambjorn:2004qm}
\bibinfo{author}{\bibfnamefont{J.}~\bibnamefont{Ambjorn}},
  \bibinfo{author}{\bibfnamefont{J.}~\bibnamefont{Jurkiewicz}},
  \bibnamefont{and} \bibinfo{author}{\bibfnamefont{R.}~\bibnamefont{Loll}},
  \bibinfo{journal}{Phys.Rev.Lett.} \textbf{\bibinfo{volume}{93}},
  \bibinfo{pages}{131301} (\bibinfo{year}{2004}), \eprint{hep-th/0404156}.

\end{thebibliography}

 \end{document}